\documentclass[twocolumn,showpacs,showkeys,preprintnumbers,amsmath,amssymb]{revtex4}

\usepackage{graphicx}
\usepackage{dcolumn}
\usepackage{bm}

\def\bfgrk #1{\mbox{\boldmath$#1$}}

\begin{document}

\title{Three Classes of Newtonian Three-Body Planar Periodic Orbits} 


\author{Milovan \v Suvakov}
\email{suki@ipb.ac.rs}
\author{V. Dmitra\v sinovi\' c}
\email{dmitrasin@ipb.ac.rs}

\affiliation{Institute of Physics Belgrade, University of Belgrade, Pregrevica 118, 11080 Beograd, Serbia }








\begin{abstract}
We present the results of a numerical search for periodic orbits of three
equal masses moving in a plane
under the influence of Newtonian gravity, with zero angular momentum. A
topological method is used to
classify periodic three-body orbits into families, which fall into four
classes, with all three previously
known families belonging to one class. The classes are defined by the
orbits’ geometric and algebraic
symmetries. In each class we present a few orbits’ initial conditions, 15 in
all; 13 of these correspond to
distinct orbits.

\end{abstract}


\maketitle   


After Bruns showed that there are 18 degrees-of-freedom, but only 10 integrals-of-motion 
in the dynamics of three Newtonian bodies, late in the 19th century, Ref. \cite{Bruns1887}, 
it has been clear that the three-body problem can not be solved in the same sense 
as the two-body one. That realization led to Poincar\'{e}'s famous {\it dictum} \cite{Poincare1899}
``... what makes these (periodic) solutions so precious to us, is that 
they are, so to say, the only opening through which we can try to penetrate 
in a place which, up to now, was supposed to be inaccessible''.
Consequently (new) periodic three-body solutions have been sought ever since, 
though a significant number were found only after 1975. 
They may be classified in three families: 
1) the Lagrange-Euler one, dating back to the eighteenth century analytical 
solutions, supplemented by one recent orbit due to Moore \cite{Moore1993},
2) the Broucke-Henon-Hadjidemetriou family, dating to the mid-1970s
\cite{Broucke1975,Hadjidemetriou1975b,Hadjidemetriou1975c,Broucke1975b,Henon1976,Henon:1977}
with periodic rediscoveries of certain members of this family
\cite{Moore1993,Vanderbei:2004}, and 3) the Figure-8 family discovered by Moore in 1993, 
Ref. \cite{Moore1993}, rediscovered in 2000, Ref. \cite{Chenciner2000}, and extended 
to the rotating case in Refs.
\cite{Nauenberg2001,Chenciner2002,Simo2002,Chenciner2005,Broucke2006,Nauenberg2007},
(see also Ref. \cite{Suvakov:2012b} and the gallery of orbits in Ref.
\cite{gallery}).

The aforementioned rediscoveries raise the issue of proper identification and 
classification of periodic three-body trajectories. Moore \cite{Moore1993} used 
braids drawn out by the three particles' trajectories in 2+1 dimensional space-time, 
Ref. \cite{Prasolov1997}, to label periodic solutions. 
This method does not associate a periodic orbit with a single braid, however, but 
with the ``conjugacy class'' of a braid group element, i.e., with all cyclic 
permutations of the strand crossings 
constituting a particular braid.
While reasonably effective for the identification of individual orbits, 
braids are less efficient at classifying orbits into families.

Montgomery \cite{Montgomery1998} suggested using the topological
properties of trajectories on the so-called shape-space sphere \cite{Iwai1987a}
to classify families of three-body orbits. That method
led Chenciner and Montgomery to their rediscovery of the figure-8 orbit \cite{Chenciner2000} and informed the present study. 
No solutions belonging to new topological classes ``higher'' than the 
figure-8 one have been found in Newtonian gravity since then, however.

Here we report the results of our ongoing numerical search for periodic collisionless
planar solutions with zero-angular-momentum in a two-parameter subspace of (the full 
four-dimensional space of) scaled zero-angular momentum initial conditions. 
This subspace is defined as that of collinear configurations with one 
body exactly in the middle between the other two, with vanishing angular momentum 
and vanishing time derivative of the hyper-radius at the initial time. 
At first we found around 50 different regions containing candidates for periodic orbits, 
at return proximity of $10^{-1}$ in the phase space, in this section of the 
initial conditions space. Then, we refined these initial conditions to the 
level of return proximity of less then $< 10^{-6}$ by using the gradient 
descent method. Here we present 15 solutions, which can be 
classified into 13 topologically distinct families. This is because two pairs of 
initial conditions specify only two independent solutions, as the respective 
members of the pairs are related by a simple rescaling 
of space and time.
Before describing these orbits and their families we must specify 
the topological classification method more closely.
\begin{figure*}[tbp]
\centerline{\includegraphics[width=7in,,keepaspectratio]{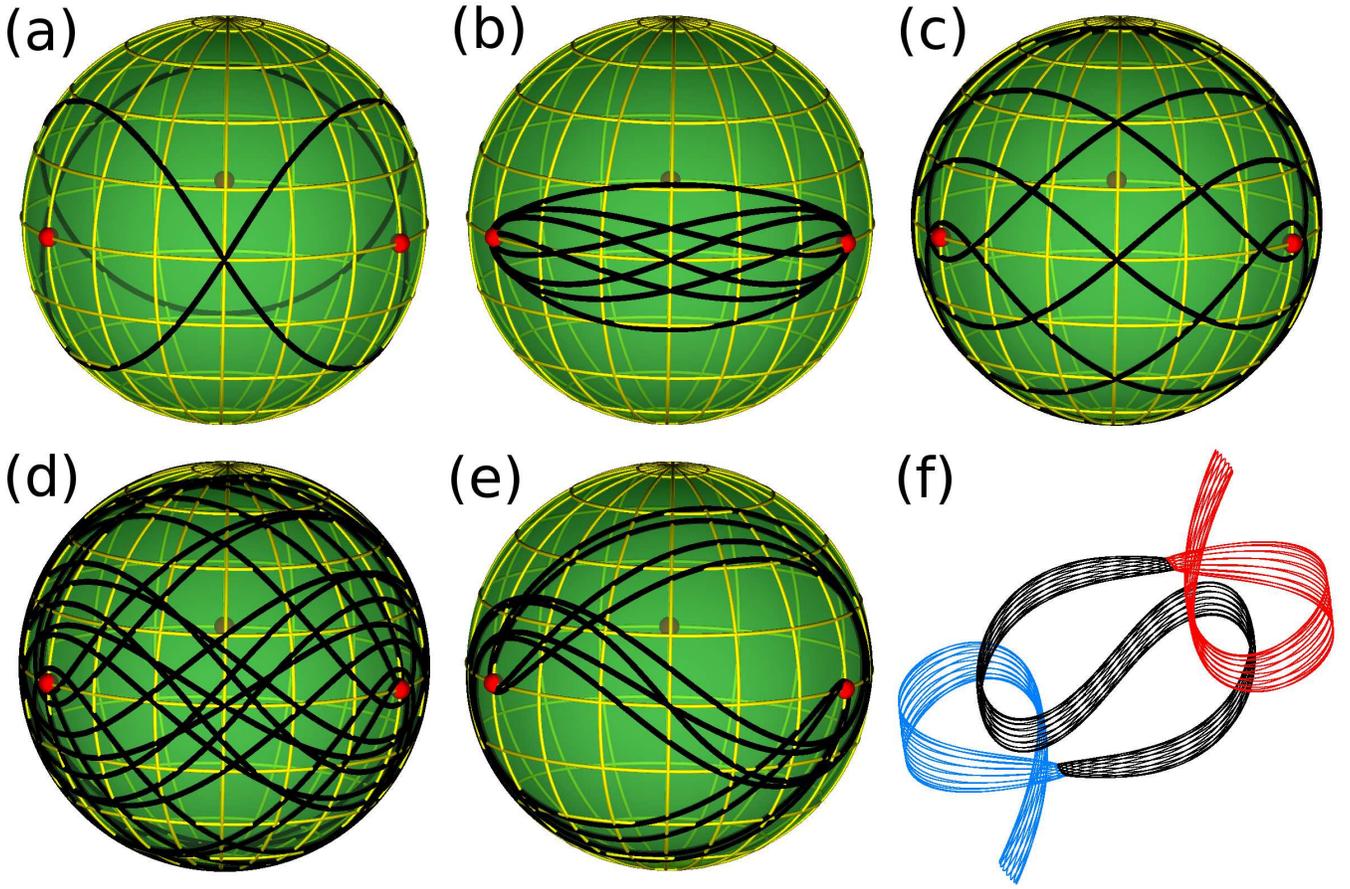}}
\caption{The (translucent) shape-space sphere, with its back side also visible 
here. Three two-body collision points (bold red circles) - punctures in the sphere - 
lie on the equator.
(a) The solid black line encircling the shape sphere twice is the figure-8 orbit.
(b) Class I.A butterfly I orbit (I.A.1). Note the two reflection symmetry axes.
(c) Class I.B moth I orbit (I.B.1) on the shape-space sphere. Note the two reflection symmetry axes.
(d) Class II.B yarn orbit (II.B.1) on the shape-space sphere. Note the single-point reflection symmetry.
(e) Class II.C yin-yang I orbit (II.C.2) on the shape-space sphere. Note the single-point reflection symmetry.
(f) An illustration of a real space orbit, the ``yin-yang II" orbit (II.C.3a). }
\label{f:sphere}
\end{figure*}


Montgomery \cite{Montgomery1998} noticed the connection between the 
``fundamental group of a two sphere with three punctures,'' i.e., 
the ``free group on two letters'' $(a,b)$, and the conjugacy classes 
of the ``projective coloured or pure braid group'' of three strands $PB_3$. 
Graphically, this method amounts to classifying closed curves according to 
their topologies on a sphere with three punctures. A stereographic projection of
this sphere onto a plane, using one of the punctures as the ``north pole'' 
effectively removes that puncture to infinity, and reduces the problem to one 
of classifying closed curves in a plane with two punctures. That leads to the 
aforementioned free group on two letters $(a,b)$, where (for definiteness) 
$a$ denotes a clockwise full turn around the right-hand-side puncture, 
and $b$ denotes the counterclockwise full turn around the other puncture, see Ref. 
\cite{Suvakov:2012b}. 
For better legibility we denote their inverses by capitalized letters $a^{-1}=A$, 
$b^{-1}=B$. Each family of orbits is associated with the conjugacy class of 
a free group element. For example the conjugacy class of the free group element 
$aB$ contains $A(aB)a = Ba$. 
To appreciate the utility of this classification one must first identify the 
two-sphere with three punctures with the shape-space sphere 
and the three two-body collision points with the punctures. 


With two three-body Jacobi relative coordinate vectors,
${\bfgrk \rho} = \frac{1}{\sqrt{2}}({\bf x_1} - {\bf x_2})$, 
${\bfgrk \lambda} = \frac{1}{\sqrt{6}}({\bf x_1} + {\bf x_2} - 
2 {\bf x_3})$,
there are three independent scalar three-body variables, i.e., 
${\bm \rho} \cdot {\bm \lambda}$, ${\bm \rho}^2$, and ${\bm
\lambda}^2$. Thus the ``internal configuration space'' of the 
planar three-body problem is three-dimensional. The hyper-radius 
$R = \sqrt{\rho^{2} + \lambda^{2}}$ 
defines the overall size of the system and removes one of 
the three linear combinations of scalar variables. Thus,
one may relate the three scalars to a (unit) hyperspace 
three-vector ${\hat {\bf n}}$ with the Cartesian components 
$n_x^{'} = \frac{2 {\bm \rho} \cdot {\bm \lambda}}{R^2}$, 
$n_y^{'} = \frac{2 {\bm \rho} \times {\bm \lambda}}{R^2}$ 
and $n_z^{'} = \frac{\lambda^2 - \rho^2}{R^2}$. 
The domain of these three-body variables is a sphere with
unit radius \cite{Iwai1987a}, see Ref. \cite{Suvakov:2012b} 
and Fig. \ref{f:sphere}(a). The equatorial circle corresponds
to collinear configurations (degenerate triangles) and the 
three points on it correspond to the two-body collisions 
(these are Montgomery's ``punctures'').  

\begin{table*}[tbh]
\begin{center} 
\caption{Initial conditions and periods of three-body orbits.
${\dot x}_{1}(0), {\dot y}_{1}(0)$ are the first particle's initial 
velocities in the $x$ and $y$ directions, respectively, $T$ is the period. 
The other two particles' initial conditions are specified by these two parameters, 
as follows,  
$x_1(0)=-x_2(0)=-1$, $x_3(0)=0$, $y_1(0)=y_2(0)=y_3(0)=0$,  
$\dot x_2(0)=\dot x_1(0)$, $\dot x_3(0)=-2\dot x_1(0)$, $\dot y_2(0)=\dot y_1(0)$,
$\dot y_3(0)=-2\dot y_1(0)$. The Newton's gravity coupling constant $G$ is taken as 
$G=1$ and equal masses as $m_{1,2,3}=1$. 
All solutions have ``inversion partners'' (mirror images) in all four quadrants, i.e. if 
${\dot x}_{1}(0), {\dot y}_{1}(0)$ is a solution, so are $\pm {\dot x}_{1}(0), \pm {\dot y}_{1}(0)$.
Some of these partners are exactly identical to the originals, others are identical up
to time reversal, and yet others are related to the originals by a reflection; we
consider all of them to be physically equivalent to the originals.
Note that two pairs of initial conditions in the same quadrant (II.C.2a and II.C.2b; 
and II.C.3a and II.C.3b) specify only two independent solutions; see 
the text for explanation.}
\begin{tabular}{lc@{\hskip 0.1in}c@{\hskip 0.1in}c@{\hskip 0.1in}l}
\hline \hline 
\setlength
{\rm Class, number and name} & ${\dot x}_{1}(0)$ & ${\dot y}_{1}(0)$ & ${\rm T}$ & {\rm Free group element} \\
\hline
\hline
I.A.1 butterfly I & 0.30689 & 0.12551 & 6.2356 & $(ab)^2 (AB)^2$ \\ 
I.A.2 butterfly II & 0.39295 & 0.09758 & 7.0039 & $(ab)^2 (AB)^2$ \\ 
I.A.3 bumblebee & 0.18428 & 0.58719 & 63.5345 & $(b^2 (ABab)^2 A^2 (baBA)^2 ba) (B^2 (abAB)^2 a^2 (BAba)^2 BA)$  \\ 
\hline
I.B.1 moth I & 0.46444 & 0.39606 & 14.8939 & $ba(BAB)ab(ABA)$ \\ 
I.B.2 moth II & 0.43917 & 0.45297 & 28.6703 &  $(abAB)^2 A (baBA)^2 B$  \\ 
I.B.3 butterfly III & 0.40592 & 0.23016 & 13.8658 & $(ab)^2(ABA)(ba)^2(BAB)$ \\ 
I.B.4 moth III & 0.38344 & 0.37736 & 25.8406 & $(babABA)^2 a (abaBAB)^2 b$ \\ 
I.B.5 goggles & 0.08330 & 0.12789 & 10.4668 &  $(ab)^2 ABBA (ba)^2 BAAB$ \\ 
I.B.6 butterfly IV & 0.350112 & 0.07934 & 79.4759 & $((ab)^2 (AB)^2)^6 A((ba)^2 (BA)^2)^6 B$ \\ 
I.B.7 dragonfly & 0.08058 & 0.58884 & 21.2710 & $(b^2(ABabAB))(a^2(BAbaBA))$ \\ 
\hline
II.B.1 yarn & 0.55906 & 0.34919 & 55.5018 & $(babABabaBA)^3$ \\ 
II.C.2a yin-yang I & 0.51394 & 0.30474 & 17.3284 & $(ab)^2(ABA)ba(BAB)$ \\ 
II.C.2b yin-yang I & 0.28270 & 0.32721 & 10.9626 & $(ab)^2(ABA)ba(BAB)$ \\ 
II.C.3a yin-yang II & 0.41682  & 0.33033 & 55.7898 & $(abaBAB)^3 (abaBAbab)(ABAbab)^3 (AB)^2$  \\ 
II.C.3b yin-yang II & 0.41734  & 0.31310 & 54.2076 & $(abaBAB)^3 (abaBAbab)(ABAbab)^3 (AB)^2$  \\ 
\hline
\hline
\end{tabular}
\label{tab:drcc2n}
\end{center}
\end{table*}

If one disallows collisions in a periodic orbit, then the orbit's trajectory 
on the sphere cannot be continuously stretched over any one of these three 
punctures, and the orbit's characteristic conjugacy class is thereby fixed; 
in this sense the topology characterizes the orbit.
Thus, periodic solutions belonging to a single collisionless family are topologically 
equivalent closed curves on the shape-space sphere with three punctures in it. 
For example, the three previously known families of orbits in shape space are shown in
Ref. \cite{Suvakov:2012b}.


One may divide the orbits into two types according to their symmetries in the shape space: 
(I) those with reflection symmetries about two orthogonal axes - the equator
and the zeroth meridian passing through the ``far'' collision point; and 
(II) those with a central reflection symmetry about a single point - the intersection
of the equator and the aforementioned zeroth meridian.
Similarly, one may divide the orbits according to algebraic exchange symmetries of 
(conjugacy classes of) their free group elements: (A) with free group elements that are symmetric under 
$a \leftrightarrow A$ and $b \leftrightarrow B$, 
(B) with free group elements symmetric under $a \leftrightarrow b$ and $A \leftrightarrow
B$, and
(C) with free group elements that are not symmetric under either of the two symmetries
(A) or (B).
We have observed empirically that, for all presently known orbits, the algebraic symmetry class 
(A) always corresponds to the geometric class (I), and that the algebraic class
(C) always corresponds 
to the geometric class (II), whereas the algebraic class (B) may  fall into either of the two geometric classes.
The first examples, to our knowledge, of higher topology trajectories on the
shape-space sphere are the two (new) zero-angular-momentum periodic solutions reported in 
Ref. \cite{Suvakov:2010}, albeit in a different (the so-called Y-string) potential. Here
we show only the new orbits in Newtonian gravity.


\noindent{\bf (I.A)} As new members of this class, we present 
three orbits in Table \ref{tab:drcc2n}: butterflies I \& II and the bumblebee. 
We show the butterfly I in Fig. \ref{f:sphere}(b). 
The butterfly's free group element is $(ab)^2 (AB)^2$. Note its close relation to the figure-8 
orbit's free group element $(ab)(AB)$ - both orbits belong to this class.
We have found two distinct butterfly orbits with the same topology (see Table \ref{tab:drcc2n})
but with different periods and sizes of trajectories, both on the shape sphere and in 
real space, see Ref. \cite{Suvakov:2012b}. 
This kind of multiplicity of solutions is not the first one of its kind: there are two 
(very similar in appearance, yet distinct) kinds of figure-8, \cite{Simo2002}. 

\noindent{\bf (I.B)} An example of this class of solutions is the 
moth I orbit, shown in Fig. \ref{f:sphere}(c). 
We have found a number of other solutions that 
belong to this class of solutions with visibly different geometrical patterns
on the shape sphere and different free group elements; see Table
\ref{tab:drcc2n} and Ref. \cite{Suvakov:2012b}.


\noindent{\bf (II.B)} An example of this class of solutions with algebraic symmetry (B), 
but with only a central geometric symmetry, is the yarn orbit (II.B.1), shown in 
Fig. \ref{f:sphere}(d). 

\noindent{\bf (II.C)} An example of this class without algebraic symmetries
is the simplest zero-angular-momentum yin-yang I orbit (II.C.2), 
shown in Fig. \ref{f:sphere}(e).  
There are two different sets of initial conditions (see Table
\ref{tab:drcc2n}) that lead 
to the same yin-yang orbit in shape space, due to the fact that this trajectory crosses 
the initial configuration on the shape-space sphere twice in one period, albeit with different 
velocities. 
Therefore the two sets of initial conditions have different energies, so that their periods 
are different, yet both correspond to the same orbit, {\it modulo} rescaling of the 
space and time, see Ref. \cite{Broucke1975b}. 
We have found four sets of initial conditions (see Table  \ref{tab:drcc2n}) 
corresponding to two distinct (i.e. having different free group elements) solutions
that belong to this (yin-yang) general class.
All yin-yang orbits seem to emerge from a single quasi-one-dimensional periodic 
orbit with collisions \cite{Suvakov:2012b},
very much like the Broucke-Henon-Hadjidemetriou ones emerge from the Schubart
(colliding) orbit \cite{Schubart1956}.

In conclusion, we have shown 13 new, distinct equal mass, zero-angular-momentum, 
planar collision-less periodic three-body orbits that can be classified in 
three new (and one old) classes. If the figure-8 orbit and its family can be 
used as a benchmark, then we expect each of the new orbits to define a 
family of periodic solutions with nonzero angular momentum. We expect our solutions 
to be either stable or marginally unstable, as otherwise they probably would not 
have been found by the present method. 

No three objects with equal masses and zero angular momentum, have been found by observational
astronomers, as yet, so our solutions cannot be directly compared with observed data
\cite{footnote1}.
Most of the three-body systems identified in observations thus far belong either to the 
(Euler-)Lagrange class, or to the quasi-Keplerian
Broucke-Hadjidemetriou-Henon class of solutions.

Besides obvious questions, such as the study of stability, and the search for the associated 
nonzero-angular-momentum solutions, there are other directions for future research, such as the 
nonequal mass solutions \cite{Galán2002}, the general-relativistic extensions of these orbits 
Ref. \cite{Imai:2007gn}, as well as the gravitational wave patterns that they generate 
\cite{Chiba:2006ad,Torigoe:2009bw}. 

Having searched in only one two-dimensional section of the full four-dimensional
space of initial conditions, we expect other types of orbits to appear, 
[single members of which have already been seen e.g. the goggles and the dragonfly 
orbits, in Ref. \cite{Suvakov:2012b}], in different sections of the full space of initial conditions.
Last, but not least, new numerical solutions in the Newtonian potential may lead to new 
analytical solutions, for example of the kind found in Ref. \cite{Fujiwara2003a} after 
the numerical discovery of the figure-8 orbit, albeit in the $-1/r^2$ potential. 
Thus, our work may shed further light on the three-body problem. 


This work was supported by the Serbian Ministry of Education, Science and
Technological Development under Grant No. OI 171037 and No. III 41011. 

\end{document}